\documentclass[epjCONF]{svjour}
\usepackage{savesym}
\usepackage{amsmath}
\savesymbol{iint}
\usepackage{graphicx}
\usepackage[varg]{txfonts} 
\usepackage[latin1]{inputenc}
\usepackage{braket}
\usepackage{bm}
\usepackage{float}
\session-title{International Conference for Atomic Physics 2012}
\begin{document}

\newcommand{\n}{\mathcal{N}}
\newcommand{\e}{\mathcal{E}}
\newcommand{\eff}{_\mathrm{eff}}
\renewcommand{\c}{\mathcal{C}}
\title{Advanced Cold Molecule Electron EDM}
\author{Wesley C. Campbell\inst{2} \and Cheong Chan\inst{1} \and David DeMille\inst{1} \and John M. Doyle\inst{3}\fnmsep\thanks{\email{doyle@physics.harvard.edu}} \and Gerald Gabrielse\inst{3} \and Yulia V. Gurevich\inst{1} \and Paul W. Hess\inst{3} \and Nicholas R. Hutzler\inst{3} \and Emil Kirilov\inst{4} \and Brendon O'Leary\inst{1} \and Elizabeth S. Petrik\inst{3} \and Ben Spaun\inst{3} \and Amar C. Vutha\inst{5}}
\institute{Yale University Department of Physics, New Haven, CT, USA \and University of California Department of Physics and Astronomy, Los Angeles, CA, USA \and Harvard University Department of Physics, Cambridge, MA, USA \and Universit\"{a}t Innsbruck Institut f\"{u}r Experimentalphysik, Innsbruck, Austria \and York University Department of Physics and Astronomy, Toronto, ON, Canada}
\abstract{
Measurement of a non-zero electric dipole moment (EDM) of the electron within a few orders of magnitude of the current best limit of $\left | d_e \right | < 1.05 \times 10^{-27} e \cdot \mathrm{cm}$ \cite{hudson2011} would be an indication of physics beyond the Standard Model. The ACME Collaboration is searching for an electron EDM by performing a precision measurement of electron spin precession in the metastable $H$ $\mathrm{{}^3\Delta_1}$ state of thorium monoxide (ThO) using a slow, cryogenic beam. We discuss the current status of the experiment. Based on a data set acquired from 14 hours of running time over a period of 2 days, we have achieved a 1-sigma statistical uncertainty of $\delta d_e = 1 \times 10^{-28}$ $e\cdot \mathrm {cm}/\sqrt{T}$, where $T$ is the running time in days.
}

\maketitle
\section{Introduction}
\label{intro}

At accelerators such as the Large Hadron Collider (LHC), particles of the highest accessible energies are used to probe physics at its most fundamental level. On a complementary front, the precise measurement techniques of atomic physics can access the vacuum fluctuations these massive particles produce. Because the search for the electron electric dipole moment (EDM) is a sensitive probe of new physics, this effort has long been at the forefront of such research \cite{khriplovich1997}\cite{bernreuther1991}. A high-precision measurement that discovers the electron EDM or sets a stringent new limit upon its size would place strong constraints on extensions to the Standard Model of particle physics (SM). A general feature of SM extensions is the prediction of an EDM for electrons and nucleons, with many theories indicating an electron EDM just below the current upper limit \cite{roberts2009}\cite{fortson2003} ($d_e < 1.05 \times 10^{-27}$ $e \cdot \mathrm{cm}$ with 90\% confidence \cite{hudson2011}, measured by the Hinds group). The symmetries of the SM, on the other hand, strongly suppress EDMs, giving rise to electron EDM predictions over a hundred billion times smaller than the current limit \cite{pospelov1991}. One well motivated SM extension is supersymmetry. Supersymmetric models require fine tuning of supersymmetric parameters to fit the current EDM limits \cite{abel2001}\cite{nir1999}. An electron EDM measurement that is 10--100 times as sensitive as the current upper bound must either observe an EDM, revealing a breakdown of the Standard Model, or set a new limit requiring such unnatural suppression of supersymmetric parameters that many supersymmetric models would have to be revised or rejected \cite{pospelov2005}.

The Advanced Cold Molecule EDM Experiment (ACME) \cite{vutha2010} is a new effort to measure the electron EDM using thorium monoxide (ThO). ThO is a polar molecule with two valence electrons. In the H $\mathrm{{}^3 \Delta_1}$ state \cite{meyer2008}, one of these electrons occupies a $\sigma$-orbital, and its EDM is relativistically enhanced due to the Sandars effect \cite{commins2007}, while the other valence electron occupies a $\delta$-orbital and allows the molecule to be easily polarized. The $\sigma$-state electron interacts with approximately 20 full atomic units of effective electric field ($\sim 100$ GV/cm) in a molecular state that can be oriented with very modest laboratory fields ($\sim 10$ V/cm) \cite{vutha2011}. The interaction of this effective molecular field with a non-zero electron EDM would manifest itself as a phase shift in ACME's Ramsey-type measurement protocol. Taking advantage of recent improvements in technologies and methods, including a new slow, cold, and intense beam source \cite{hutzler2011} and ThO's near-ideal $\mathrm{{}^3\Delta_1}$ state structure (see e.g. \cite{vutha2010}\cite{meyer2006}\cite{vuthathesis2011}), we have developed an experiment with the unprecedented electron EDM statistical sensitivity of about $1 \times 10^{-28}$ $e \cdot \mathrm{cm}$ in one day of averaging time. This is 10 times better than the current experimental limit \cite{hudson2011}. As discussed below, ACME's systematic errors are also projected to be smaller than those of past experiments and can be checked with high precision on the time scale of days. We are currently studying various possible sources of systematic error in preparation for reporting a new result. 

\section{Atomic and molecular electron EDM experiments}
\label{sec:amoedm}

The signature of a permanent electron EDM, $d_e$, is an energy shift $\varepsilon_\mathrm{EDM}$ of an unpaired electron (or electrons) in an electric field $\bm{\mathcal{E}}$: 
\begin{equation} \label{eq:Hedm}
\varepsilon_{\mathrm{EDM}} = -\vec{d}_e \cdot \bm{\e}. 
\end{equation}
In the vicinity of some atomic nuclei, electrons experience very strong electric fields \cite{commins2007}\cite{salpeter1958}\cite{sandars1965}. These internal atomic and molecular fields can be partially or completely oriented by polarizing the atom or molecule, which together with relativistic effects gives the electron EDM a non-zero average energy shift. Per Eq. (\ref{eq:Hedm}), this shift can be interpreted as an interaction between $d_e$ and an average effective electric field $\e \eff$ produced by the atomic nucleus. The size of $\e \eff$ can be shown to scale approximately as the cube of the atomic number $Z$ \cite{budker2004}. Thus, the species that yield the most sensitive (i.e. largest $\varepsilon_{\mathrm{EDM}}$) electron EDM measurements are heavy (large $Z$), highly polarizable atoms and molecules with unpaired valence electrons whose wavefunctions have a large amplitude near the nucleus.

These principles have guided the search for electron EDM for the last fifty years, during which time the strongest limits have consistently been set by atomic and molecular experiments. Table \ref{tab:statcomparison} summarizes the two most recent EDM upper bounds, obtained with atomic thallium (Tl) and the polar molecule ytterbium fluoride (YbF), and compares the sensitivity of these experiments with ACME's demonstrated sensitivity.

\begin{table}
\centering
\caption{Comparison of the statistical sensitivity of ACME with that of the two experiments placing the strongest limits on the electron EDM. The estimated statistical uncertainty for the Tl and YbF experiments assumes a duty cycle of 100\%.}
\label{tab:statcomparison}
\begin{tabular}{lcccc}
\hline\noalign{\smallskip}
Experiment & Species & Statistical Uncertainty & Upper Limit on & References \\
& &  After 1 Day of Averaging ($e \cdot \mathrm{cm}$) & $|d_e|$ ($e \cdot \mathrm{cm}$) & \\
\noalign{\smallskip}\hline\noalign{\smallskip}
Hinds et al. & YbF & $\sim 2 \times 10^{-27}$ & $1.05 \times 10^{-27}$ & \cite{hudson2011} \cite{sauer2011} \cite{kara2012} \\
\hline\noalign{\smallskip}
Commins et al. & Tl & $\sim 5 \times 10^{-28}$ & $1.7 \times 10^{-27}$ & \cite{regan2002} \cite{commins1994} \\
\hline\noalign{\smallskip}
ACME & ThO & $\sim 1 \times 10^{-28}$ & Experiment in progress & \cite{vutha2010} \\
\hline\noalign{\smallskip}
\end{tabular}
\end{table}

\subsection{Thorium monoxide electron EDM}
\label{sec:ThO}

ACME's molecule of choice, ThO, combines the aforementioned benefits of a high-Z, polar molecule with several other powerful advantages. These properties of ThO conspire to increase ACME's statistical sensitivity compared to previous electron EDM experiments, mitigate the technical demands of working with molecules rather than atoms, and suppress or rule out many systematic errors \cite{vutha2010}.

\begin{figure}
\begin{minipage}[t]{0.66\linewidth}
\centering
\includegraphics[width=\textwidth]{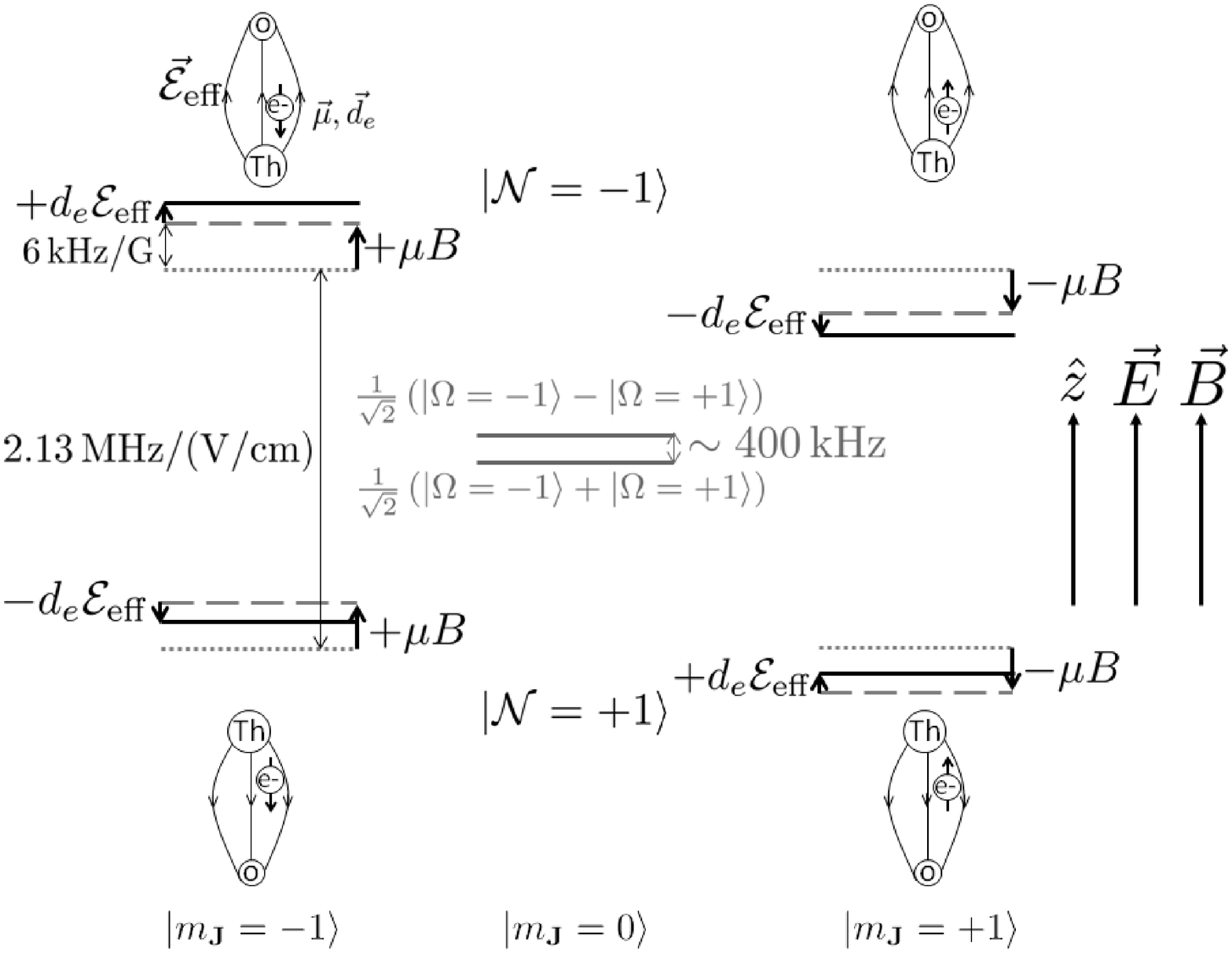}
\caption{Sublevel structure of the $H$ $\mathrm{{}^{3}\Delta_{1}}$ state of ThO. In the absence of applied $\vec{E} \parallel \hat{z}$ and $\vec{B} \parallel \hat{z}$ fields, the stationary states are the $\Omega$-doubled parity eigenstates $\frac{1}{\sqrt{2}}(\ket{\Omega = +1} \pm \ket{\Omega = -1})$, which are split by a few hundred kHz (solid gray lines). $\vec{E}$-fields of $\sim 10$ V/cm fully mix these doublets in the $M_J \equiv \hat{z} \cdot \vec{J} = \pm 1$ states by resolving the aligned and anti-aligned orientations ($\n \equiv \mathrm{sgn}(\hat{n} \cdot \vec{E}) = \mathrm{sgn}(\hat{z} \cdot \vec{E}) M_J \Omega = \pm 1$) of the internuclear axis $\hat{n}$. The linear Stark splitting between these $\mathcal{N}$ states (dotted gray lines) is measured to be 2.13 MHz/(V/cm). In an applied $\vec{B}$-field, the measured Zeeman shift (dashed gray lines) between the $M_J=\pm 1$ states of each $\n$ sublevel is $\pm$12 kHz/G \cite{vutha2011}. If $d_e \neq 0$, these $M_J$ levels experience an additional relative shift equal to $\pm 2 d_e \e \eff$ (solid black lines). These relative shifts are in opposite directions in the two $\mathcal{N}$ levels since $\bm{\e}\eff$ points in opposite directions. The ACME experiment is performed by measuring the energy shift between the states $\ket{\mathcal{N},M_{J} = -1}$ and $\ket{\mathcal{N},M_J = +1}$ for both $\mathcal{N}$ as a function of the electric and magnetic field and looking for a shift that depends only on the signs of $\mathcal{N}$ and $\vec{E}$. See Sections \ref{sec:measurement} and \ref{sec:analysis}.}
\label{fig:H-state}
\end{minipage}
\hspace{0.3cm}
\begin{minipage}[t]{0.31\linewidth}
\centering
\includegraphics[width=\textwidth]{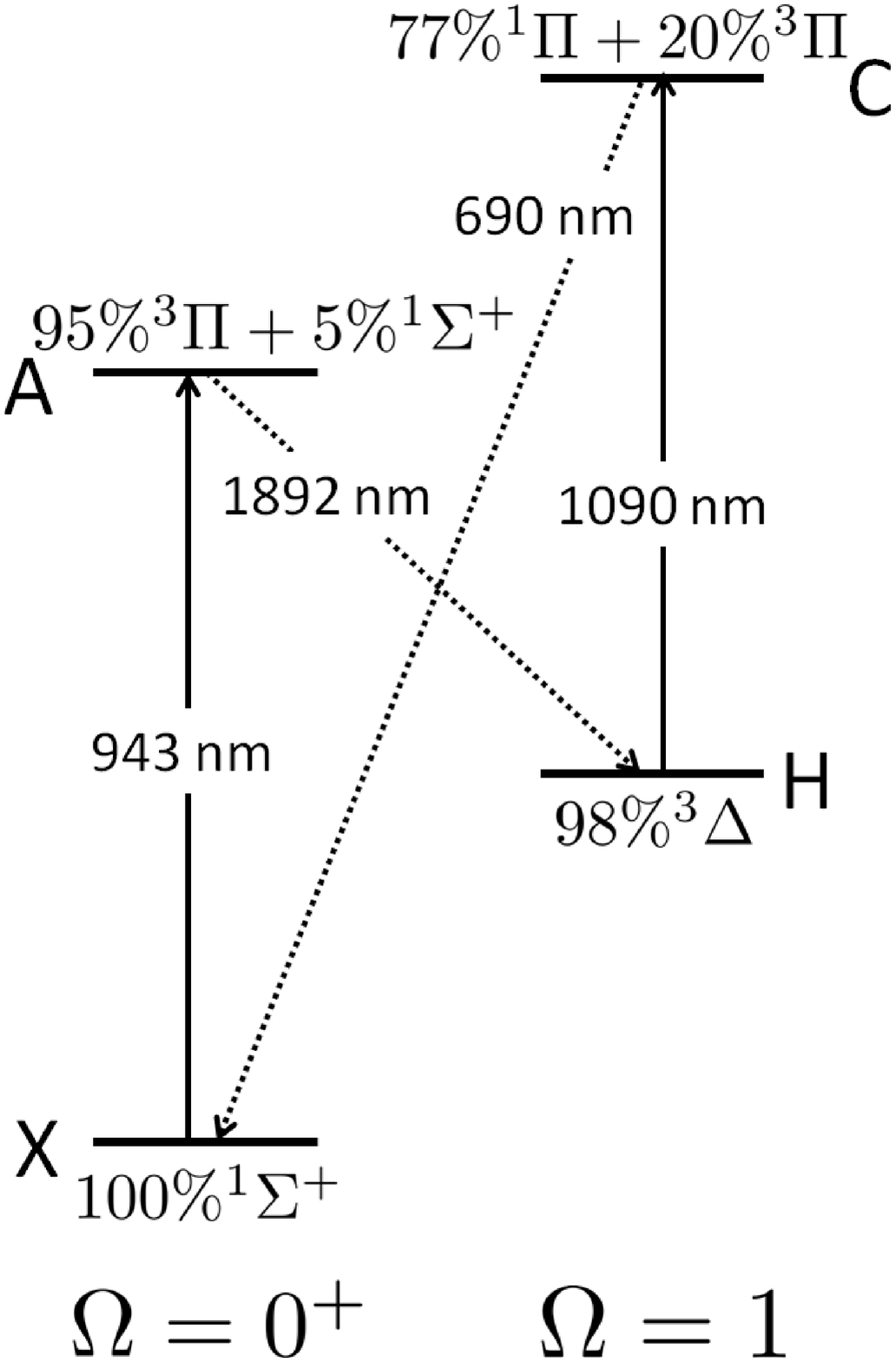}
\caption{Key levels and transitions in ThO, based on \cite{vutha2010}\cite{edvinsson1985}\cite{paulovic2003}. All relevant states are in the ground vibrational level. The electronic states are denoted by letters, and the angular momentum character of each state is indicated by molecular spectroscopy symbols. The wavelength of each transition is given in nm. The ACME measurement scheme makes use of both diode laser pumped excitations (solid arrows), and spontaneous decays (dotted arrows), as described in Section \ref{sec:measurement}.}
\label{fig:ThOlevels}
\end{minipage}
\end{figure}

Meyer and Bohn \cite{meyer2008} have calculated the effective internal electric field $\e \eff$ of fully polarized ThO to be $\sim 100$ GV/cm, which is among the largest of any investigated species. This field is nearly 4 times as large as the estimated field in fully polarized YbF \cite{mosyagin1998}, nearly 8 times as large as the $\e \eff$ achieved in partially polarized YbF in the Hinds experiment \cite{sauer2011}, and over 1000 times larger than the $\e \eff$ achieved in the Tl experiment  \cite{regan2002}. Moreover, ThO possesses a low-lying metastable state $H$ $\mathrm{{}^3 \Delta_1}$ (see Fig. \ref{fig:H-state}), which exhibits several features beneficial to an EDM experiment. Firstly, it has a measured lifetime of 1.8 ms \cite{vutha2010}, sufficient to perform our Ramsey experiment in a molecular beam with a coherence time of 1.1 ms (see Section \ref{sec:measurement}). This is comparable to the coherence times in both the YbF (642 $\mu$s \cite{hudson2011}) and the Tl ($\sim$2.5 ms \cite{reganthesis2001}) electron EDM experiments. Secondly, the spin and orbital magnetic moments of a state with $\mathrm{{}^3 \Delta_1}$ angular momentum cancel almost perfectly \cite{vutha2010}, and the residual g-factor is measured to be $g_{H,J=1} = 4.3(3) \times 10^{-3}$ \cite{vutha2011}.\footnote{To avoid confusion with similar definitions of the molecular g-factor, we specify that in the present paper's notation, the energy shift of a Zeeman sublevel of $H$, $J=1$ in an applied magnetic field is given by $\varepsilon_B = g_{H,J=1} \mu_B \vec{J} \cdot \vec{B}$.}  This small magnetic moment renders the experiment highly insensitive to magnetic field imperfections.

Finally, the most advantageous property of the $H$ $\mathrm{{}^3 \Delta_1}$ state of ThO is its extremely large static electric dipole polarizability resulting from a pair of nearly degenerate, opposite-parity sublevels split by only a few hundred kHz \cite{meyer2008}\cite{edvinsson1984}\cite{meyer2006}. This level structure gives polarizabilities on the order of $10^4$ or more times larger than for a more typical diatomic molecule state, in which an applied electric field polarizes the molecule by mixing opposite-parity rotational levels typically spaced by many GHz. The opposite-parity sublevels $H$, $J=1$ state are formed by even and odd combinations of molecular orbitals with opposite signs of the quantum number $\Omega \equiv \hat{n} \cdot \vec{J}$ (the projection of the total angular momentum on the molecular bond axis) and are a general feature of states with $\Omega \geq 1$ in Hund's case (c) molecules \cite{herzberg1950}\cite{demille2000}. Such ``$\Omega$-doubled'' states are immensely valuable to electron EDM searches because they can be fully mixed in electric fields of only a few tens or hundreds of V/cm, completely polarizing the molecule \cite{demille2000}\cite{demille2001}. Thus, EDM experiments on molecules with $\Omega$-doublets can take full advantage of the molecules' effective internal field while avoiding the technical challenges and potential systematic errors introduced by large lab fields. Furthermore, because the effective electric field in a fully polarized molecule is independent of the externally applied electric field $\vec{E}$, the electron EDM signal is also independent of the magnitude of the applied field [see Eq. (\ref{eq:phi})], allowing such experiments to set limits on systematic effects correlated with $|E|$. Another benefit of the $\Omega$-doublet in ThO is that the polarized $H$-state molecule can be spectroscopically prepared with its dipole either aligned or anti-aligned with $\vec{E}$, allowing us to switch the sign of the electric field experienced by the electron EDM without physically changing the laboratory field \cite{kawall2004}. As discussed in Section \ref{sec:systematic}, this provides a way to rule out systematic errors correlated with the sign of the applied field, such as leakage currents, motional magnetic fields, and geometric phases \cite{vutha2010}\cite{vutha2009}. The ACME experiment is currently taking data to improve its statistics and set limits on possible systematic errors. 

Besides these features, ThO also provides manifold technical advantages. All of the relevant optical transitions (see Fig. \ref{fig:ThOlevels}) are well studied \cite{edvinsson1984}\cite{huber1979}\cite{marian1987}\cite{watanabe1997}\cite{paulovic2002}\cite{goncharov2005} and accessible to diode lasers. In addition, ThO has no nuclear spin and so avoids the complexities of hyperfine structure. Finally, despite the fact that ThO is chemically reactive and its precursors are highly refractory, it can be produced in large quantities in a cryogenic buffer gas beam \cite{hutzler2011} (see Section \ref{sec:ThObeam}).

\section{ACME experiment overview}
\label{sec:ACMEexp}

In order to measure the electron EDM, ACME produces a high-flux beam of ThO and uses an optical state preparation and readout scheme to detect the Ramsey fringe phase shift resulting from a non-zero $\vec{d}_e \cdot \bm{\e} \eff$. The measurement and apparatus are described here.

\subsection{Measurement scheme}
\label{sec:measurement}

\begin{figure}
\centering
\resizebox{1.\columnwidth}{!}{\includegraphics{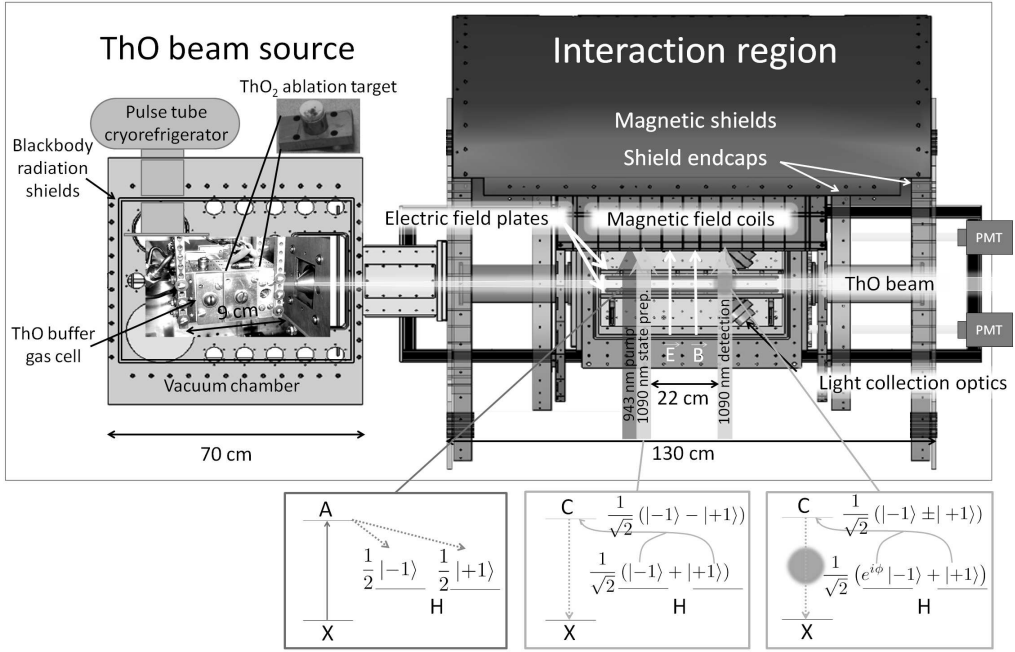}[h]}
\caption{Schematic of the ACME apparatus and measurement described in Section \ref{sec:ACMEexp}. On the left, a pulse of gas-phase ThO molecules is produced and cooled in a buffer gas cell and flows out towards the right in a beam (see Section \ref{sec:ThObeam}). This beam enters a magnetically shielded interaction region where uniform, parallel $\vec{E}$- and $\vec{B}$-fields are applied. At the entrance of the field region, the molecules are pumped from the $\ket{X, v=0, J=1, M_J = \pm 1}$ states to the $\ket{A, v=0, J=0, M_J = 0}$ state, where they spontaneously decay to the $\ket{H, v=0, J=1}$ state, equally populating the $\ket{J=1,M=\pm 1, \n = \pm 1}$ sublevels (see Fig. \ref{fig:H-state}). Next, a pure superposition of Zeeman sublevels $\ket{X_\n}$ [see Eq. (\ref{eq:XY})] is prepared in one of the two $\mathcal{N}$ states of H by pumping out the orthogonal superposition $\ket{Y_\n}$ using linearly polarized light resonant with the $\ket{H, v=0, J=1, \mathcal{N}= \pm 1} \rightarrow \ket{C, v=0, J=1, M_J=0}$ transition. Next, the molecule state precesses in the applied $\vec{E}$ and $\vec{B}$ fields for approximately 1.1 ms as the beam traverses the 22-cm-long interaction region. The relative phase accumulated between the Zeeman sublevels depends on $d_e$ through Eq. (\ref{eq:phi}). Near the exit of the field region, we read out the final state of the molecules: By exciting the $\ket{H, v=0, J=1, \mathcal{N}= \pm 1} \rightarrow \ket{C, v=0, J=1, M_J=0}$ transition with rapidly switched orthogonal ($\hat{x}$ and $\hat{y}$) linear polarizations and detecting the $C \rightsquigarrow X$ fluorescence from each polarization, we project the population onto the $\ket{X_\n}$ and $\ket{Y_\n}$ states. The phase from Eq.~(\ref{eq:phi}) is given by $\cos 2 \phi = \mathcal{A}$ [see Eq. (\ref{eq:asymmetry})].}
\label{fig:experiment_schematic}
\end{figure}

The ACME apparatus and measurement scheme are illustrated in Fig. \ref{fig:experiment_schematic} and described in \cite{vutha2010}. Molecules from the beam source enter the interaction region and are intercepted by an optical pumping laser tuned to the $X \to A$ transition (see Fig. \ref{fig:ThOlevels}). Excitation by this laser and subsequent $A \rightsquigarrow H$ spontaneous decay populate the $H$ state. The measurement is performed in select sublevels in the ground ro-vibrational level ($v=0,J=1$) of the $H$ state. In the absence of an applied electric field $\vec{E}$, sublevels in this manifold are identified by their quantum numbers $M_J = \pm 1, 0$ (projection of $\mathbf{J}$ along the lab-frame quantization axis $\hat{z}$), and $P = \pm 1$ (parity). The opposite-parity $\Omega$-doublet levels in the $H$ state have a very small splitting ($\sim$ 400 kHz \cite{meyer2008}\cite{vutha2011}\cite{edvinsson1984}), which we neglect. When a sufficiently large (more than $\sim 10$ V/cm) electric field $\vec{E}$ is applied collinear with $\hat{z}$, the $P = \pm 1$ sublevels with the same value of $M_J$ mix completely; the resulting eigenstates have complete electrical polarization, described by the quantum number $\mathcal{N} \equiv \mathrm{sgn}\left ( \hat{n} \cdot \vec{E} \right ) = \pm 1$. (The $M_J=0$ sublevels do not mix.)  The relevant energy levels are shown in Fig. \ref{fig:H-state}. The tensor Stark shift $\mathrm{\Delta}_{\mathrm{St}}$ is defined as the magnitude of the shift of the oriented $|M_J| = 1$ levels from the unperturbed $M_J=0$ levels. A magnetic field $\vec{B} \approx 10$ mG is also applied collinear with $\hat{z}$, lifting the degeneracy of the $M_J = \pm 1$ levels.

Since the $H$ state is populated by spontaneous decay from $A$, it is initially in a mixed state, with all sublevels used in the experiment approximately equally populated. By coupling the molecules to a strong state-preparation laser driving the $H \to C$ transition, we deplete the coherent superposition of $|M_J=\pm 1; \mathcal{N}\rangle$ that couples to the laser polarization $\hat{\epsilon}_p$, leaving behind a dark state. With the laser polarization $\hat{\epsilon}_p = \hat{y}$ for example, the prepared state of the molecules is 
\begin{equation}\label{eq:psii}
\ket{\psi_i^\n} = \frac{1}{\sqrt{2}} \left ( \ket{M_J = +1;\n} + \ket{M_J=-1;\n} \right )
\end{equation}
with $\mathcal{N}=+1$ ($\mathcal{N}=-1$) corresponding to the lower (upper) $\Omega$-doublet component. The tensor Stark shift $\mathrm{\Delta}_{\mathrm{St}}$ is large enough that levels with different values of $\mathcal{N}$ are spectrally resolved by the state preparation laser. Hence a particular value of $\mathcal{N}$ is chosen by appropriate tuning of the laser frequency.

The molecules in the beam then travel through the interaction region, where the relative phase of the two states in the superposition is shifted by the interaction of $\bm{\mu}_{H,J=1}$ with $\vec{B}$ and $\mathbf{d}_e$ with $\bm{\e} \eff$. The energy shifts of the $M_J=\pm 1$ levels in Fig. \ref{fig:H-state} are given approximately by
\begin{equation}\label{eq:H-state-energies}
\varepsilon(M_J,\n,\vec{E},\vec{B}) = g_{H,J=1}M_J\mu_B B \hat{B} - d_{H,J=1}\n E +d_e\e_{\mathrm{eff}} M_J\n \hat{E},
\end{equation}
where $g_{H,J=1} = 4.3(3)\times 10^{-3}$ and $d_{H,J=1} = 0.84(2) \; ea_0$ are the magnetic g-factor and electric dipole moments of the $H,J=1$ state respectively \cite{vutha2011}, $\mu_B$ is the Bohr magneton, $e$ is the electron charge, and $a_0$ is the Bohr radius. The terms (from left to right) give the interaction of the magnetic dipole with the external magnetic field, the Stark shift $\mathrm{\Delta}_\mathrm{St}$, and the interaction of the electron EDM with the effective molecular field. Here we assume that the $H$-state is fully polarized, which occurs in external fields of $\sim 10$ V/cm, much smaller than the typical experimental field of 140 V/cm. The magnitudes of applied field vectors are given in Roman font, e.g. $B = |\vec{B}|$. The hat denotes the sign of a quantity's projection on the lab-fixed quantization axis of the experiment, e.g. $\hat{B} = \textrm{sgn} (\hat{z} \cdot \vec{B})$. This simple formula neglects a large number of important terms, such as the electric field dependence of the g-factors \cite{bickman2009}, background fields, motional fields, etc., but this expression will be sufficient to explain the basic measurement procedure.

After free evolution during flight (over a distance $L = 22$ cm in our experiment), the final wavefunction of the molecules is
\begin{equation}\label{eq:psif}
\ket{\psi_f^\n} = \frac{1}{\sqrt{2}} \left (e^{i\phi} \ket{M_J=+1;\n} + e^{-i\phi}\ket{M_J=-1;\n} \right ).
\end{equation}
For a molecule with velocity $v$ along the beam axis, the accumulated phase $\phi$ can be expressed as 
\begin{align}
\phi & = \int_{x=0}^{x=L} \left [ \varepsilon(M_J = +1,\n, \vec{E}, \vec{B}) - \varepsilon(M_J = -1,\n, \vec{E}, \vec{B}) \right ] \frac{dx}{2 \hbar v} \\ 
& = \int_{x=0}^{x=L} \left ( d_e \e \eff \n \hat{E} + g_{H,J=1} \mu_B B \hat{B} \right ) \frac{dx}{\hbar v} \equiv \phi_{\e} + \phi_{B}. \label{eq:phi}
\end{align} 

Using the fact that our beam source has a narrow forward velocity distribution (with average forward velocity $v$ and spread $\mathrm{\Delta} v_\parallel \ll v$, see Section \ref{sec:ThObeam}), we make the approximation that all molecules experience the same phase shift as they traverse the interaction region. Furthermore, because the $\vec{E}$- and $\vec{B}$-fields are highly uniform along the length of the interaction region, we can pull out the integrand and write:
\begin{align}
\phi_\e & \approx d_e \e\eff \n \hat{E} \frac{L}{\hbar v}, \: \textrm{and} \label{eq:phiE} \\
\phi_B & \approx g_{H,J=1} \mu_B B \hat{B} \frac{L}{\hbar v} \label{eq:phiB}
\end{align}
for all molecules in the beam.

The phase $\phi$ is detected by measuring populations in two ``quadrature components'' $\ket{X_\mathcal{N}}$ and $\ket{Y_\mathcal{N}}$ of the final state, where we define
\begin{eqnarray}
|X_\mathcal{N}\rangle & \equiv & \frac{1}{\sqrt{2}} \left ( \ket{M_J=+1; \n} + \ket{M_J=-1; \n} \right ), \: \mathrm{and} \nonumber \\ 
|Y_\mathcal{N}\rangle & \equiv & \frac{1}{\sqrt{2}} \left ( \ket{M_J=+1; \n} - \ket{M_J=-1;\ \n} \right ).  \label{eq:XY}
\end{eqnarray}
The quadrature state $\ket{X_\n}  (\ket{Y_\n})$ is independently detected by excitation with a laser coupling the $H$ and $C$ states whose polarization is $\hat{\epsilon}_{d} = \hat{x}$ ($\hat{\epsilon}_{d} = \hat{y}$). The $C$ state quickly decays to the ground state, emitting fluorescence at 690 nm, which we collect with an array of lenses and focus into fiber bundles and light pipes. These in turn deliver the light to two photomultiplier tubes (PMT's),\footnote{Hamamatsu R8900U-20} where it is detected. This scheme allows for efficient rejection of scattered light from the detection laser since the emitted fluorescence photons are at a much shorter wavelength than the laser.

The probability of detecting a molecule in the quadrature state $\ket{X_\n}$ ($\ket{Y_\n}$), given by $P_{X} = |\braket{ X_\n|\psi_f^\n}|^2$ ($P_{Y} = |\braket{ Y_\n|\psi_f^\n}|^2$), can be expressed as $P_{X}  =  \cos^2 \phi$ ($P_{Y} = \sin^2 \phi$).  The detected fluorescence signal from each quadrature state is proportional to its population.  We express these signals ($S_X$ and $S_Y$) as a number of photoelectron counts per beam pulse, and write $S_{X(Y)} = S_0 P_{X(Y)}$, where $S_{0}$ is the total signal from one beam pulse. Thus, $S_X$ and $S_Y$ trace out two sinusoidal curves (or Ramsey fringes) of opposite phase as a function of applied magnetic field. For the highest sensitivity to $d_e$, we ``sit on the side of the Ramsey fringe'' where small changes in $\phi_\e$ are most noticeable, i.e. where $\partial/\partial\phi_\e [S_{X(Y)}]$ is maximized. Therefore, we adjust the magnetic field to yield a bias phase $|\phi_B| = \pi/4$ and rewrite $S_X$ and $S_Y$ as
\begin{equation}
S_X \approx S_0 \left ( - \hat{B} \phi_\e + \frac{1}{2} \right ), \: \textrm{and} \: S_Y \approx S_0 \left ( + \hat{B} \phi_\e + \frac{1}{2} \right ). \\
\end{equation}
Then the EDM phase $\phi_\e$ can be determined by constructing the quantity $\mathcal{A}$, known as the asymmetry:
\begin{align}
\mathcal{A} & \equiv \frac{S_Y - S_X}{S_X + S_Y} \approx 2 \hat{B} \phi_\e \label{eq:asymmetry} \\
\frac{\hat{B}}{2}\mathcal{A} & \approx \phi_\e \label{eq:measphiE}.
\end{align}
Note from Eq. (\ref{eq:phiE}) that $\phi_\e$ is odd in $\vec{E}$ and $\n$, even in $\vec{B}$, and proportional to $\e\eff$. In Section \ref{sec:analysis} we discuss how to use these correlations to isolate the EDM term from various systematic effects. 

The shot-noise limited statistical uncertainty in $\phi_\e$ is $1/(2 \c \sqrt{N})$, where $N$ is the total number of photon counts and the quantity $\c$ introduced in this expressions is the Ramsey fringe contrast (or visibility), which accounts for inefficiencies in state preparation and varying precession times for different molecules. Therefore, the shot-noise limited uncertainty in the measured EDM value is [from differentiating $d_e$ with respect to $\phi_\e$ in Eq. (\ref{eq:phiE})] \cite{vutha2010}
\begin{equation} \label{eq:EDMuncert}
\delta d_e = \frac{\hbar}{2 \c \tau \e\eff(\dot{N}T)^{1/2}},
\end{equation}
where $\tau=L/v$ is the precession time of the molecules in the fields, $\dot{N}$ is the time-averaged counting rate of the detectors, and $T$ is the total experimental running time. The quantities $\tau$ and $\e \eff$ are determined by physical properties of the $H$-state, as described above, and the large ThO fluxes achieved by the ACME beam source help to keep our uncertainty low by providing large $\dot{N}$.

\subsection{ThO buffer gas beam}
\label{sec:ThObeam}

ACME uses a cryogenic buffer gas beam source to achieve high single-quantum-state intensities of the chemically unstable molecular species ThO. The heart of the cold beam apparatus, the buffer gas cell (see Fig.~\ref{fig:experiment_schematic}), is similar to those described in earlier buffer gas cooled beam publications \cite{maxwell2005}\cite{patterson2007}\cite{patterson2009}\cite{hutzler2012}. Our ACME beam was carefully characterized and described in \cite{hutzler2011}. The cell is a small copper chamber mounted in vacuum and held at a temperature of 16 K with a Cryomech PT415 pulse tube cooler. Cold neon buffer gas flows into the cell through a fill line at one end of the cylindrical volume, and at the other end of the cell, an aperture 5 mm in diameter in a thin (0.5 mm) plate is open to the external vacuum, allowing the buffer gas to flow out as a beam. The cell is surrounded by two nested chambers of metal that are also thermally anchored to the pulse tube cooler. The inner chamber is held at 4 K and acts as a high-speed, large-capacity cryopump for neon, maintaining a high vacuum of $\sim 3$ $\mathrm{\muup}$Torr in the system despite large buffer gas throughputs. The outer chamber is kept at 50 K and serves to shield the inner cryogenic regions from blackbody radiation emitted by the room temperature vacuum chamber. Both the 4 K and the 50 K chambers have a window to admit the ablation laser and apertures to transmit and collimate the buffer gas beam.

The source of ThO molecules is a ceramic target of thoria (ThO${}_2$) made in-house using established techniques \cite{balakrishna1988}\cite{vutha2010}. ThO molecules are introduced into the cell via laser ablation: A Litron Nano TRL 80-200 pulsed Nd:YAG laser is fired at the ThO${}_2$ target, creating an initially hot plume of gas-phase ThO molecules. The ablation pulse energy is set to 75--100 mJ and the repetition rate to 50 Hz. On a time scale rapid compared to the emptying time of the cell into the beam region, the hot ThO molecules thermalize with the 16 K buffer gas in the cell. Continuous neon flow at $\sim 40$ SCCM (standard cubic centimeters per minute) maintains a buffer gas density of $n_0 \approx 10^{15}$--$10^{16}$ cm${}^{-3}$ ($\approx 10^{-3}$--$10^{-2}$ Torr, where the subscript ``$0$'' indicates the steady-state value of the quantity in the cell). This is sufficient for rapid translational and rotational thermalization of the molecules and for producing hydrodynamic flow out of the cell aperture that entrains a significant fraction of the molecules before they can diffuse to the cell walls and stick. The result is a 1--3 ms long pulsed beam of cold ThO molecules embedded in a continuous flow of buffer gas.

Just outside the cell exit, the buffer gas density is still high enough for ThO--Ne collisions to play a significant role in the beam dynamics. The average thermal velocity of the buffer gas atoms is higher than that of the molecules by a factor of $\sqrt{m_\mathrm{mol}/m_\mathrm{b}}$, where the subscripts ``$\mathrm{b}$'' and ``$\mathrm{mol}$'' indicate buffer gas and molecule quantities, respectively. Consequently, the ThO molecules ($m_\mathrm{mol}=248$ amu) experience collisions primarily from behind, with the fast neon atoms ($m_\mathrm{b}=20$ amu) pushing the slower ThO molecules ahead of them as they exit the cell. This accelerates the molecules to an average forward velocity $v_f$ that is larger than the thermal velocity of ThO. As the buffer gas pressure in the cell is increased, $v_f$ approaches $v_{0,\mathrm{b}}$, the thermal velocity of the buffer gas. 

The angular distribution of a beam has a characteristic apex angle $\theta$ given by $\tan({\theta/2}) \equiv \mathrm{\Delta} v_{\perp}/2 v_{f}$, where $\mathrm{\Delta} v_{\perp}$ is the transverse velocity spread of the beam. For the ACME beam, the apex angle is $\theta \approx 30^{\circ}$, and the characteristic solid angle is $\Omega \approx 0.3$ sr. The beam velocity is measured to be $\sim 180$ m/s. As the gas cloud expands nearly isentropically out of the cell into the vacuum, it must also cool. The measured final longitudinal and rotational temperature of the beam is $\sim 4$ K, yielding a forward velocity distribution $\mathrm{\Delta} v_{\parallel}$ of $\sim 30$ m/s FWHM (full width at half maximum) and efficiently populating low-lying rotational levels in the ground electronic state (e.g. $\sim 30\%$ in $J=1$). The total number of molecules per pulse in the few most populated quantum states is measured to be $\sim 10^{11}$. This slow, cold, high-intensity molecular beam provides ACME with a long interaction time $\tau$ over a short distance, low phase decoherence due to the narrow velocity spread, and a high count rate $\dot{N}$.

\section{Data analysis}
\label{sec:analysis}

\begin{figure}
\centering
\resizebox{0.57\columnwidth}{!}{\includegraphics{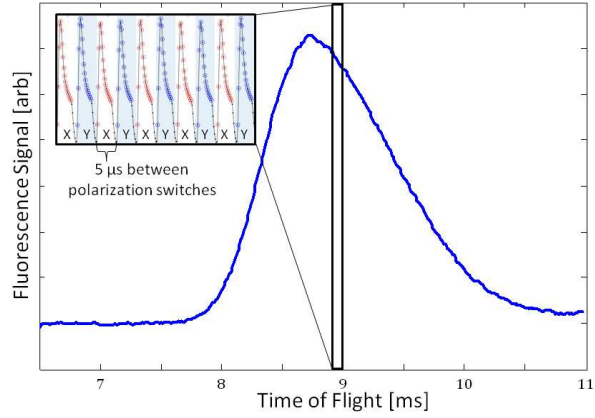}}
\caption{Average fluorescence signal from a molecule pulse vs. time since ablation. Each pulse of molecules is $\sim 2$ ms wide. The inset shows a zoom-in on the fluorescence signal over a 50 $\mu$s interval, revealing the 100 kHz chopping of the probe laser polarization between $\hat{\epsilon}_d = \hat{x}$ and $\hat{\epsilon}_d = \hat{y}$, used to measure $S_X$ and $S_Y$, respectively.}
\label{fig:SignalPulse}
\end{figure}

Figure \ref{fig:SignalPulse} shows some example data collected using the scheme described in Section \ref{sec:ACMEexp}. As derived in Section \ref{sec:measurement}, this measurement scheme determines the accumulated phase due to the energy shift between the two $M_J$ levels in either $\n$ state. This energy shift is given by [see Eq. (\ref{eq:H-state-energies})]:
\begin{eqnarray}\label{eq:H-state-deltaE}
\mathrm{\Delta} \varepsilon(\n,\vec{E},\vec{B}) & \equiv & \varepsilon(M_J=+1,\n, \vec{E}, \vec{B}) - \varepsilon(M_J=-1,\n, \vec{E}, \vec{B})\\
& = & 2g_{H,J=1}\mu_B B \hat{B} + 2d_e\e_{\mathrm{eff}} \n\hat{E}
\end{eqnarray}
If we wish to measure $d_e$ in a way that is insensitive to noise or uncertainty in the external magnetic field $\vec{B}$, we can repeat the measurement with both $\pm \vec{B}$ and take the sum of the measurements, $\mathrm{\Delta} \varepsilon(\n, \vec{E}, \vec{B}) + \mathrm{\Delta} \varepsilon(\n, \vec{E},-\vec{B}) = 4 d_e\e_{\mathrm{eff}} \n\hat{E}$.  We can then take the difference of the measurements to isolate the magnetic field interaction, $\mathrm{\Delta} \varepsilon(\n, \vec{E}, \vec{B}) - \mathrm{\Delta} \varepsilon(\n, \vec{E}, -\vec{B}) = 4g_{H,J=1}\mu_B B$.  In other words, since the spin precession in the magnetic field is ``$\vec{B}$-odd" (reverses when $\vec{B}$ is reversed), and the electron EDM precession is ``$\vec{B}$-even", we can distinguish them by taking repeated measurements with reversing magnetic fields and looking at sums or differences of those measurements. Notice that we can also separate the spin and EDM precession by reversing $\n$ or $\vec{E}$ since the two terms also have opposite parity under reversal of those quantities.

In a real experiment a number of uncontrolled effects are present, including background fields, correlated fields (e.g. magnetic fields from leakage currents which reverse synchronously with $\vec{E}$), motional fields, geometric phases, and many more \cite{khriplovich1997}. Despite the best experimental efforts, these effects may cause energy shifts larger than the electron EDM; however, we can isolate the electron EDM from these effects using its unique  ``$\n \vec{E} \vec{B} = --+$" parity, i.e. odd parity under molecular dipole or electric field reversal and even parity under magnetic field reversal.

If we perform 8 repeated experiments, with each of the $2^3 = 8$ combinations of $\pm\n,\pm \vec{E},\pm \vec{B}$, we can take sums and differences to compute the 8 different possible parities under $\n, \vec{E}, \vec{B}$ reversals, as shown in Table \ref{tab:parity-sums}. Apart from higher-order terms, such as cross-terms between background electric and magnetic fields, the electron EDM is the only term with $\n \vec{E} \vec{B} = --+$ parity. This technique of isolation by parity is how EDM experiments can perform sensitive measurements of the electron EDM with achievable levels of control of experimental parameters. We also perform a number of auxiliary switches to check for other systematic dependences of the $\n \vec{E} \vec{B} = --+$ signal, such as rotating the polarization angle of the pump and probe lasers and interchanging the positive and negative field plate voltage leads.

\subsection{Statistical sensitivity}
\label{sec:statistical}

The shot-noise limited sensitivity of the ACME experiment is given by Eq. (\ref{eq:EDMuncert}). Other sources of technical noise may cause the achieved experimental sensitivity to be larger, but our measurements indicate that we are very near the shot noise limit \cite{kirilov_unpub}.

\begin{table}
\centering
\caption{Shot-noise limited electron EDM uncertainty estimated from measured and calculated quantities. The measured uncertainty is about $1.4$ times the shot noise limit. Quantities in bold are ingredients in Eq. (\ref{eq:EDMuncert}). All quantities other than the effective electric field $\e_\mathrm{eff}$ are either experimental inputs or are derived from measurements taken in the ACME experiment's ordinary running configuration as described in the text.}
\label{tab:sensitivity}
\begin{tabular}{lll}
\hline\noalign{\smallskip}
Quantity & Value & Formula \\
\noalign{\smallskip}\hline\noalign{\smallskip}
\textbf{Effective electric field} & $\mathbf{104 \pm 26}$ \textbf{GV/cm} \cite{meyer2008} & $\bm{\e}_{\mathrm{eff}}$ \\
\hline\noalign{\smallskip}
\textbf{Interaction time} & $\mathbf{1.1 \pm 0.1}$ \textbf{ms} & $\bm{\tau}$ \\ 
\hline\noalign{\smallskip}
\textbf{Contrast} & $\mathbf{90 \pm 5 \%}$ & $\bm{\mathcal{C}}$ \\ 
\hline\noalign{\smallskip}
$\:\:\:\:$ Molecule beam brightness & & \\
$\:\:\:\:$ per quantum state per pulse & 6--18 $\times 10^{10}$ sr${}^{-1}$ & \\ 
$\; \times$ Solid angle subtended by detection region & $6.3 \pm 0.6 \times 10^{-5}$ sr & \\ 
$\; \times$ Pulse rate & $50$ Hz & \\
$\; \times$ Molecule fraction in EDM state & $4.1 \pm 0.8 \times 10^{-2}$ & \\ 
$\; \times$ Detection efficiency & $1.0 \pm 0.2 \times 10^{-2}$ & \\ 
$\; \times$ Duty cycle & $0.5 \pm 0.1 $ & \\ 
\textbf{Count rate (calculated from above)} & \textbf{3--14} $\mathbf{\times 10^4}$ \textbf{s}$\mathbf{{}^{-1}}$ & $\mathbf{\dot N}$\\
\textbf{Count rate (directly measured)} & \textbf{$\bm{\sim}$5} $\mathbf{\times 10^4}$ \textbf{s}$\mathbf{{}^{-1}}$ & $\mathbf{\dot N}$\\
\noalign{\smallskip}\hline\noalign{\smallskip}
EDM uncertainty in a total running time of $T$ & & $\delta d_e = \frac{\hbar}{ 2 \mathcal{C} \mathcal{E}_\mathrm{eff} \tau \sqrt{\dot N T}}$ \\
$\:\:\:\:$ From calculated $\dot{N}$ & 2--9 $\times 10^{-29}\; e \cdot \mathrm{cm} \sqrt{\mathrm{days}/T}$ &  \\
$\:\:\:\:$ From measured $\dot{N}$ & $\sim$6 $\times 10^{-29}\; e \cdot \mathrm{cm} \sqrt{\mathrm{days}/T}$ &  \\
\noalign{\smallskip}\hline
\end{tabular}
\end{table}
Table \ref{tab:sensitivity} derives ACME's expected shot-noise limited statistical EDM sensitivity from measured and calculated quantities. In this table, the interaction time $\tau$ is equal to the length of the interaction region $L = 22$ cm divided by the measured beam velocity $v = 180$ m/s \cite{hutzler2011}. The contrast $\mathcal{C}$ is determined by measuring the slope of the Ramsey fringe at $|\phi_B| = \pi/4$. 

The count rate can be determined directly, by converting the PMT signal to a photon number, or indirectly, by starting with the measured molecule beam intensity and multiplying by the efficiency of each step in the measurement scheme. The molecule beam brightness in a single $M_J$ sublevel of $\ket{X, J=1}$ was reported in \cite{hutzler2011}, and the solid angle of the molecular beam used in the measurement is given by geometry: The final molecular beam collimator is 1 cm $\times$ 1 cm in area and is 126 cm from the beam source, so $\Omega_\mathrm{detect} = (1$ cm)${}^2/(126$ cm$){}^2$. The pulse rate of the YAG is set to 50 Hz. The fraction of molecules available for detection is given by:
\begin{align}
\textrm{Mol. fraction in EDM state} & = (\textrm{optical pumping efficiency of X $\to$ A $\rightsquigarrow$ H}) \nonumber \\
\times ( & \textrm{fraction of H state sublevels used}) \times \exp[{-\tau/(\textrm{H state lifetime})}] \nonumber \\
\times ( & \textrm{Beam attenuation due to background collisions}) \label{eq:molsurvive} \\
& = 0.67 \times 1/6 \times \exp (-1.2\;\mathrm{ms}/1.8\;\mathrm{ms}) \times 0.8 = 0.04,
\end{align}
where each value in Eq. (\ref{eq:molsurvive}) was measured separately. The fluorescence detection efficiency is the product of the measured geometric collection efficiency of the detection optics ($\sim 14\%$) and the quantum efficiency of the PMT's ($10\%$). The duty cycle is the fraction of the time during the run that data is being collected. ACME's duty cycle is presently around 50\% because of the time required to switch various parameters (e.g. laser polarization angle), degauss the magnetic shields, optimize the ablation yield, and tune up the lasers during the run.

\begin{figure}
\centering
\resizebox{0.57\columnwidth}{!}{\includegraphics{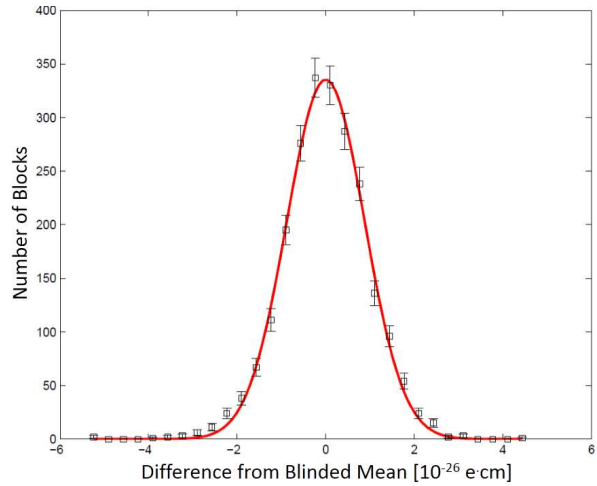}}
\caption{Distribution of blinded values of the electron EDM calculated from data taken over a total of 14 hours on 2 separate days. Each of the 2300 measurements plotted on the histogram was calculated from one ``block'' of data, where each block consists of 800 molecular beam pulses with various parameter switches. The error bars show the standard deviation on the number of blocks in each histogram bin. The solid line is a fitted Gaussian distribution with a standard deviation of $8.5 \times 10^{-27}$ $e \cdot \mathrm{cm}$.  The 1-sigma statistical uncertainty in the EDM from this plot is $1.6 \times 10^{-28}$ $e \cdot \mathrm{cm}$ in 14 hours, which corresponds to a 1-sigma statistical error bar of about $1 \times 10^{-28}$ $e \cdot \mathrm{cm}$ in one day of averaging time.}
\label{fig:EDMhistogram}
\end{figure}

Figure \ref{fig:EDMhistogram} shows a set of EDM data (with an unknown blind offset added during data processing) taken over a total of 14 hours on 2 different days. The 1-sigma statistical uncertainty in the EDM from this plot is $1.6 \times 10^{-28}$ $e \cdot \mathrm{cm}$ in 14 hours. This corresponds to a 1-sigma statistical error bar of about $1 \times 10^{-28}$ $e \cdot \mathrm{cm}$ in one day of averaging time, which is consistent within uncertainty with 1.4 times the shot-noise limit estimated in Table \ref{tab:sensitivity}.

\subsection{Systematic checks}
\label{sec:systematic}

As discussed above, the particular behavior of the electron EDM under reversal of applied electric field, applied magnetic field, and molecule electric dipole orientation allows for powerful rejection of systematic effects. In order to test our ability to reject experimental imperfections, we can purposely amplify these imperfections and study their effect on our measured electron EDM. Say that some quantity $X$ (for example, a non-reversing electric or magnetic field) mimics the electron EDM according to the relation $d_{e,\mathrm{false}}(X) = \alpha X$. If the quantity $X$ can only be determined or controlled to the level $X_{\mathrm{control}}$, then our measurement will have a systematic uncertainty due to imperfections in $X$ of order $\delta d_{e,X} \approx |\alpha X_{\mathrm{control}}|$. The quantity $X_{\mathrm{control}}$ can typically be determined with direct measurements (magnetometers to measure magnetic fields, spectroscopic techniques to measure electric fields, optical cavities to determine laser noise, etc.), but it remains to determine $\alpha$. The general technique to determine $\alpha$ is simply to measure $d_e$ with varying values of $X$ and fit the functional form of $d_{e,\mathrm{false}}(X)$. At the time of this writing, no known systematic effects in the ThO experiment, including effects due to background fields, motional fields, and geometric phases, are expected to be larger than $\sim 10^{-32}$ $e \cdot \mathrm{cm}$, well below the statistical sensitivity of the experiment in reasonable averaging time \cite{vutha2010}. Nevertheless, we are currently in the process of varying a large number of experimental parameters to look for unexpected systematic effects.

\begin{table}
\centering
\caption{Parity of energy shifts of selected effects in the ACME measurement. The difference between the g-factors of the two $\n$-states of $H$ is $\mathrm{\Delta} g$ \cite{bickman2009}, and the subscript $\mathrm{nr}$ denotes the non-reversing component of an applied field. Products of terms denote correlations between those terms. The terms with $+--$ parity are higher-order and negligibly small.}
\label{tab:parity-sums}      
\begin{tabular}{cl}
\hline\noalign{\smallskip}
$\n \vec{E} \vec{B}$ Parity & Quantities \\
\noalign{\smallskip}\hline\noalign{\smallskip}
$+++$ & Electron spin precession in background (non-reversing) magnetic field $B_\mathrm{nr}$, \\
& Pump/probe relative polarization offset \\
$++-$ &  Electron spin precession in applied magnetic field \\
$+-+$ &  Leakage currents $B_\mathrm{leak}$\\
$-++$ & $\mathrm{\Delta} g B_\mathrm{nr}$, $\mathrm{\Delta} g B_\mathrm{leak}E_\mathrm{nr}$\\
$+--$ &  --- \\
$-+-$ &  Electric-field-dependent g-factors \cite{bickman2009}\\
$--+$ &  Electron EDM\\
$---$ & $\mathrm{\Delta} g E_{\mathrm{nr}}$\\
\noalign{\smallskip}\hline
\end{tabular}
\end{table}

\section{Conclusion}
\label{sec:conclusion}

The discovery of an electron EDM or an improvement on its upper limit by an order of magnitude or more would have a significant impact on our understanding of fundamental particle physics. We have described an ongoing experiment to search for the electron EDM using cold ThO molecules. This experiment has achieved a one-sigma statistical uncertainty of $1 \times 10^{-28} e\cdot \mathrm {cm}/\sqrt{T}$, where T is the running time in days. This advance over previously published electron EDM experiments was made possible by the combination of a greatly increased molecular flux provided by our new cold molecular beam source and our choice of the ThO molecule, which is fully polarizable in small fields and has the highest effective electric field of any investigated species. We are now working to put limits on systematic errors that may be present in the experiment. ThO, due to its advantageous level structure, is particularly well suited to the suppression and rejection of systematic effects while searching for the electron EDM.


\begin{thebibliography}{}

\bibitem{hudson2011}
J. J. Hudson, D. M. Kara, I. J. Smallman, B. E. Sauer, M. R. Tarbutt, and E. A. Hinds, Nature \textbf{473}, (2011) 493--496.

\bibitem{khriplovich1997}
I. B. Khriplovich and S. K. Lamoreaux, \textit{CP Violation Without Strangeness: Electric Dipole Moments of Particles, Atoms, and Molecules} (Springer-Verlag, Berlin 1997).

\bibitem{bernreuther1991}
W. Bernreuther and M. Suzuki, Reviews of Modern Physics \textbf{63}, (1991) 313--340.

\bibitem{roberts2009}
E. D. Commins and D. P. DeMille, ``The Electric Dipole Moment of the Electron." In B. L. Roberts and W. J. Marciano, Editors, \textit{Lepton Dipole Moments} (World Scientific, Singapore 2009) 519--581.

\bibitem{fortson2003}
N. Fortson, P. Sandars, and S. Barr, Physics Today \textbf{56}, (2003) 33--39.

\bibitem{pospelov1991}
M. E. Pospelov and I. B. Khriplovich, Soviet Journal of Nuclear Physics \textbf{53}, (1991) 638--640.

\bibitem{abel2001}
S. Abel, S. Khalil, and O. Lebedev, Nuclear Physics B \textbf{606}, (2001) 151--182.

\bibitem{nir1999}
Y. Nir, ArXiv (1999) 9911321v2.

\bibitem{pospelov2005}
M. Pospelov and A. Ritz, Annals of Physics \textbf{318}, (2005) 119--169.

\bibitem{vutha2010}
A. C. Vutha, W. C. Campbell, Y. V. Gurevich, N. R. Hutzler, M. Parsons, D. Patterson, E. Petrik, B. Spaun, J. M. Doyle, G. Gabrielse, and D. DeMille, Journal of Physics B \textbf{43}, (2010) 074007.

\bibitem{meyer2008}
E. L. Meyer and J. L. Bohn, Physical Review A \textbf{78}, (2008) 01052(R).

\bibitem{commins2007}
E. D. Commins, J. D. Jackson, and D. P. DeMille, American Journal of Physics \textbf{75}, (2007) 532--536.

\bibitem{vutha2011}
A. C. Vutha, B. Spaun, Y. V. Gurevich, N. R. Hutzler, E. Kirilov, J. M. Doyle, G. Gabrielse, and D. DeMille, Physical Review A \textbf{84}, (2011) 034502.

\bibitem{hutzler2011}
N. R. Hutzler, M. F. Parsons, Y. V. Gurevich, P. W. Hess, E. Petrik, B. Spaun, A. Vutha, D. DeMille, G. Gabrielse, and J. M. Doyle, Physical Chemistry Chemical Physics \textbf{13}, (2011) 18976--18985.

\bibitem{meyer2006}
E. R. Meyer, J. L. Bohn, and M. P. Deskevich, Physical Review A \textbf{73}, (2006) 062108.

\bibitem{vuthathesis2011}
A. C. Vutha, Ph.D. Thesis, Yale University (2011).

\bibitem{sauer2011}
B. E. Sauer, J. J. Hudson, D. M. Kara, I. J. Smallman, M. R. Tarbutt, and E. A. Hinds, Physics Procedia \textbf{17}, (2011) 175--180.

\bibitem{kara2012}
D. M. Kara, I. J. Smallman, J. J. Hudson, B. E. Sauer, M. R. Tarbutt, and E. A. Hinds, New Journal of Physics \textbf{14}, (2012) 103051.

\bibitem{regan2002}
B. C. Regan, E. D. Commins, C. J. Schmidt, and D. DeMille, Physical Review Letters \textbf{88}, (2002) 071805.

\bibitem{commins1994}
E. D. Commins, S. B. Ross, D. DeMille, and B. C. Regan, Physical Review A \textbf{50}, (1994) 2960--2977.

\bibitem{salpeter1958}
E. E. Salpeter, Physical Review \textbf{112}, (1958) 1642--1648.

\bibitem{sandars1965}
P. G. H. Sandars, Physics Letters \textbf{14}, (1965) 194--196.


\bibitem{budker2004}
D. Budker, D. F. Kimball, and D. P. DeMille, \textit{Atomic Physics: An Exploration Through Problems and Solutions} (Oxford University Press, Inc., New York 2004).

\bibitem{mosyagin1998}
M. S. Mosyagin, M. G. Kozlov, and A. V. Titov, Journal of Physics B \textbf{31}, (1998) L763--L767.

\bibitem{reganthesis2001}
B. C. Regan, Ph.D. Thesis, Berkeley (2001).

\bibitem{edvinsson1984}
G. Edvinsson and A. Lagerqvist, Physica Scripta \textbf{30}, (1984) 309--320.

\bibitem{herzberg1950}
G. Herzberg, \textit{The Spectra and Structures of Simple Free Radicals} (Cornell University Press, Ithaca 1971).

\bibitem{demille2000}
D. DeMille, F. Bay, S. Bickman, D. Kawall, D. Krause, Jr., S. E. Maxwell, and L. R. Hunter, Physical Review A \textbf{61}, (2000) 052507.

\bibitem{demille2001}
D. DeMille, F. Bay, S. Bickman, D. Kawall, L. R. Hunter, D. Krause, Jr., S. Maxwell, and K. Ulmer, American Institute of Physics Conference Proceedings \textbf{596}, (2001) 72--83.

\bibitem{kawall2004}
D. Kawall, F. Bay, S. Bickman, Y. Jiang, and D. DeMille, AIP Conference Proceedings \textbf{698}, (2004) 192--195.

\bibitem{vutha2009}
A. Vutha and D. DeMille, ArXiv, (2009) 0907.5116v1.

\bibitem{edvinsson1985}
G. Edvinsson and A. Lagerqvist, Journal of Molecular Spectroscopy \textbf{113}, (1985) 93--104.

\bibitem{paulovic2003}
J. Paulovi\u{c}, T. Nakajima, K. Hirao, R. Lindh, and P. A. Malmqvist, Journal of Chemical Physics \textbf{119}, (2003) 798--805.

\bibitem{huber1979}
K. P. Huber and G. Herzberg \textit{Constants of Diatomic Molecules} (Van Nostrand Reinhold, New York 1979).

\bibitem{marian1987}
C. M. Marian, U. Wahlgren, O. Gropen, and P. Pyykko, Journal of Molecular Structure \textbf{169}, (1987) 339.

\bibitem{watanabe1997}
Y. Watanabe and O. Matsuoka, Journal of Chemical Physics \textbf{107}, (1997) 3738--3739.

\bibitem{paulovic2002}
J. Paulovi\u{c}, T. Nakajima, and K. Hirao, Journal of Chemical Physics \textbf{117}, (2002) 3597.

\bibitem{goncharov2005}
V. Goncharov, J. Han, L. A. Kaledin, and M. C. Heaven, Journal of Chemical Physics \textbf{122}, (2005) 204311.

\bibitem{bickman2009}
S. Bickman, P. Hamilton, Y. Jiang, and D. DeMille, Physical Review A \textbf{80}, (2009) 023418.

\bibitem{maxwell2005}
S. E. Maxwell, N. Brahms, R. DeCarvalho, D. R. Glenn, J. S. Helton, S. V. Nguyen, D. Patterson, J. Petricka, D. DeMille, and J. M. Doyle, Physical Review Letters \textbf{95}, (2005) 173201.

\bibitem{patterson2007}
D. Patterson and J. M. Doyle, Journal of Chemical Physics \textbf{126}, (2007) 154309.

\bibitem{patterson2009}
D. Patterson, J. Rasmussen, and J. M. Doyle, New Journal of Physics \textbf{11}, (2009) 55018.

\bibitem{hutzler2012}
N. R. Hutzler, H. Lu, and J. M. Doyle, Chemical Reviews \textbf{112}, (2012) 4803--4827.

\bibitem{balakrishna1988}
P. Balakrishna, B. P. Varma, T. S. Krishnan, T. R. R. Mohan, and P. Ramakrishnan, Journal of Materials Science Letters \textbf{7}, (1988) 657--660.

\bibitem{kirilov_unpub}
E. Kirilov et al., ``Shot noise-limited spin measurements in a pulsed molecular beam,'' in preparation.


\end{thebibliography}
\end{document}